\documentstyle[12pt]{article}
\input epsf                   
\date  {{\small Feb. 28, 1998}} 
\def\abst#1{\begin{minipage}{5.25in}
{\noindent   \normalsize
{\bf Abstract} #1}  \\ \end{minipage}  }
\makeatletter
\newcounter{refctr}
\def\Alphref#1{
\@ifundefined{r@#1}{??}
{\setcounter{refctr}{\ref{#1}}\Alph{refctr}}}
\makeatother
\newtheorem{thm}{Theorem}
\newtheorem{lem}{Lemma}
\newtheorem{cor}{Corollary}
\def\be{\begin{equation}}
\def\ee{\end{equation}}
\def\ben{\begin{displaymath}}
\def\een{\end{displaymath}}
\def\bea{\begin{eqnarray}}
\def\eea{\end{eqnarray}}
\def\bean{\begin{eqnarray*}}
\def\eean{\end{eqnarray*}}

\def\Z{\mbox{\boldmath $Z$}}
\def\R{\mbox{\boldmath $R$}}
\def\E{\mbox{\boldmath $E$}}

\def\iu{{\rm i}}
\def\e{{\rm e}}
\def\Im{{\rm Im}}

\def\Ind{{\rm Index\,}}

\def\Arg{{\rm Arg\,}}

\def\remark{\noindent {\bf Remark:}  }

\newcommand{\eq}[1]{eq.~(\ref{#1})}       

\newcounter{masectionnumber}
\newcommand{\masect}[1]{\setcounter{equation}{0}
  \refstepcounter{masectionnumber} \vspace{1truecm plus 1cm} \noindent
    {\large\bf \arabic{masectionnumber}. #1}\par \vspace{.2cm}
      \addcontentsline{toc}{section}{\arabic{masectionnumber}. #1}
    }
 \renewcommand{\theequation}
    {\mbox{\arabic{masectionnumber}.\arabic{equation}}}

    \newcounter{masubsectionnumber}[masectionnumber]
\newcommand{\masubsect}[1]{
    \refstepcounter{masubsectionnumber} \vspace{.5cm} \noindent
  {\large\em \arabic{masectionnumber}.\alph{masubsectionnumber} #1}
    \par\vspace*{.2truecm}

\addcontentsline{toc}{subsection}
 {\arabic{masectionnumber}.\alph{masubsectionnumber}\hspace{.1cm} #1}
    }
\newcommand{\startappendix}{ \setcounter{masectionnumber}{0} }
\newcommand{\maappendix}[1]{
    \setcounter{equation}{0}
  \refstepcounter{masectionnumber} \vspace{1truecm plus 1cm} \noindent
    {\large\bf Appendix \Alph{masectionnumber}. #1}\par \vspace{.2cm}
    \renewcommand{\theequation}
    {\mbox{\Alph{masectionnumber}.\arabic{equation}}}
   \addcontentsline{toc}{section}{\Alph{masectionnumber}. #1}
      }
 \newcommand{\masubsectapp}[1]{
    \refstepcounter{masubsectionnumber} \vspace{.5cm} \noindent
  {\large\em \Alph{masectionnumber}.\arabic{masubsectionnumber} #1}
    \par\vspace*{.2truecm}

\addcontentsline{toc}{subsection}
 {\arabic{masectionnumber}.\alph{masubsectionnumber}\hspace{.1cm} #1}
    }

\begin{document}

\title{\vspace*{-.35in}
Localization Bounds for an Electron Gas }
\author{M. Aizenman
${}^{(a)}$ \qquad and \qquad G.M. Graf ${}^{(b)}$\\ \hskip 1cm
\vspace*{-0.05truein} \\
\normalsize \it  ${}^{(a)}$ Departments of Physics and Mathematics,
Princeton University \\
\normalsize \it Jadwin Hall, Princeton, NJ 08544, USA \\
\normalsize \it ${}^{(b)}$ Theoretische Physik,
ETH-H\"onggerberg, CH--8093 Z\"urich, Switzerland }
\maketitle
\thispagestyle{empty}        

\abst{ Mathematical analysis of the Anderson localization has been
facilitated by the use of suitable fractional moments of the Green
function.  Related methods permit now a readily accessible derivation
of a number of physical manifestations of localization, in regimes
of strong disorder, extreme energies, or weak disorder away from the
unperturbed spectrum.  The present work establishes on this basis
exponential decay for the modulus of the two--point function, at all
temperatures as well as in the ground state,  for a Fermi gas within
the one--particle approximation.  Different implications, in particular
for the Integral Quantum Hall Effect, are reviewed.
         }
\bigskip

  \renewcommand{\baselinestretch}{1}
  {
  \renewcommand{\thefootnote}{}
  \footnotetext{
  {\copyright \ Copyright rests with the authors.  Faithful
  reproduction for non-commercial purpose is permitted.}}   }


\newpage
\masect{Introduction}
\vspace{-.6cm}
\masubsect{The localization condition}

This is a report on recent progress in the mathematical
analysis of Anderson localization.
The simplifications which have been made in its
derivation permit now to access a number of interesting properties
of systems with disorder, by methods which are both mathematically
rigorous and not excessively complicated. The report includes some
new technical statements, but we recall also a number of previously
known results, derived by various other authors, in order to present
a more complete picture of the physically motivated
questions which can be addressed by related mathematical methods.

Anderson localization was first discussed in the context of the
conduction properties of metals (\cite{An,Th74}), but the mechanism is of
relevance in a variety of other situations (e.g., \cite{FiKl}). The basic
phenomenon is that disorder can cause localization of electron states (or
normal modes --- in other systems) and thereby affect properties like: time
evolution (non--spreading of wave packets), conductivity (in response to
electric field), Hall currents (in the presence of both magnetic and
electric field), and statistics of the spacing between nearby energy levels.

In the electron gas approximation the system of electrons in a
crystal is modeled by a gas of Fermions moving on a lattice.
We focus here on systems with homogeneous disorder, which otherwise
are periodic or translation invariant, at least up to gauge
transformations.  The excitations of the system are
described by an effective one--body Hamiltonian, which consists of a
short--range hopping term and a local potential.
The one--particle
Hamiltonian is a self--adjoint operator with matrix elements of the form
\begin{equation}
H = K_{x,y} + U^{per}_x+\lambda V_x \;,
\label{eq:h}
\end{equation}
acting in the Hilbert space $\ell^2(\mbox{\boldmath
$Z$}^d)$, where $K_{x,y}$ is a short--range hopping term,
$U^{per}_x$ a periodic potential, and $\lambda V$  a
random potential expressing the disorder (impurities) with a tunable
strength parameter $\lambda$.

We shall not discuss here the validity of the one--particle
approximation, or that of the linear response theory.  Instead, we
focused on the analysis within such frameworks.  In particular, we
shall demonstrate how resolvent estimates can be used to address a
number of physically motivated questions.

For $K_{x,y}$ we consider the following two cases:

\noindent
\noindent{\em No magnetic field:}
$K_{x,y}$ depend only on the difference $(x-y)$.

\noindent{\em Constant magnetic field:}
There is some ambiguity in the definition of the magnetic flux,
since flux differences of
$h/e$ can be induced, or compensated for, by gauge
transformations. For concreteness sake, let us restrict to the
operators of the form:
\begin{equation}
K_{x,y} \ = \ {\rm e}^{-{\rm i}{\cal A}_{x,y}}\ \delta_{|x-y|,1}
\label{eq:hb}
\end{equation}
with a phase ${\cal A}_{x,y}$ which is an antisymmetric function of
the oriented bonds, $b=\{x,y\}$.
(${\cal A}_{x,y}$ can be viewed as the line integral of the
``vector potential''
$\times (-e/\hbar)$ along the direct path from $x$ to $y$).
The magnetic flux through a plaquette $P$ is taken to be
\be
B_{P}\  = \ - \frac{\hbar}{e} \sum_{b\in \partial P} \Arg(K_{b})\;,
\ee
with the argument function interpreted through
its principal branch, i.e.,  $-\pi < \Arg z \le \pi$.
At non-zero field, translation invariance
is possible only in the sense of magnetic translations, which
combine shifts with gauge transformations, i.e. are unitaries
of the form
\begin{equation}
U(a)|x\rangle\ =\ {\rm e}^{-{\rm i}\varphi_{a}(x)}|x+a\rangle \;.
\label{eq:m1}
\end{equation}
In such cases, $K=U(a)KU(a)^*$ implies ordinary
translation invariance for gauge invariant quantities, such as
$|K_{x,y}|$ and $|\langle x| (K-E-{\rm i}\eta)^{-1} |y \rangle |$.
(The fact that the composition law for the magnetic shifts
provides only a projective representation of the translation
group does not affect our analysis.)

The potential $V$ is realized as a collection of independent
identically distributed random variables $V_x$, whose probability
distribution may be of the form $r(v) dv$ with $r(v)$ a bounded
probability--density function. (These conditions may be relaxed: the
results described below are valid also for a broad class of
correlated randomness, more singular probability distributions
for $V$, and Hamiltonians with
off diagonal disorder, i.e., randomness in $K_{x,y}$.)

Of central importance in the analysis is the Green
function, i.e., the kernel of the resolvent operator:
\begin{equation} G(x,y;E+{\rm i}\eta) = \langle x|
{1 \over H-E-{\rm i}\eta}
|y\rangle  \ \ .
\end{equation}
The behavior of this function
at $\eta = 0+$ reveals a great deal about the spectral properties of
the Hamiltonian (e.g., discrete versus continuous spectrum), the
nature of its eigenfunctions (localized or extended), and the
response of the system (e.g., to electric fields) at the
linear--response level.

A technically convenient signature of localization is a bound on the
fractional moments of $G(0,x;E)$.  The explicit condition is that for
energies $E$ in an interval $[a,b]$ and some $0<s<1$,
\begin{equation}
\mbox{\boldmath $E$}( |G(x,y;E+{\rm i}\eta)|^s ) \le C^s{\rm
e}^{-s\mu |x-y|} \;,
\label{eq:gs}
\end{equation}
for all $\eta\neq0$. Here and henceforth
{\boldmath $E$} represents the average over the randomness and
$C < \infty$ and $\mu > 0$ are
constants which may change from line to line, but are to be
understood as independent of $\eta$.
The value of $s$ is of little consequence (if the condition
(\ref{eq:gs}) holds for some $s$ then  by H\"older inequality it
extends to all smaller $s>0$), but the restriction $s<1$ permits to
avoid the divergence explained below.

\begin{figure}[t]
    \begin{center}
     \leavevmode
     \hbox{ %
 	\epsfxsize=5.5in
  \epsffile{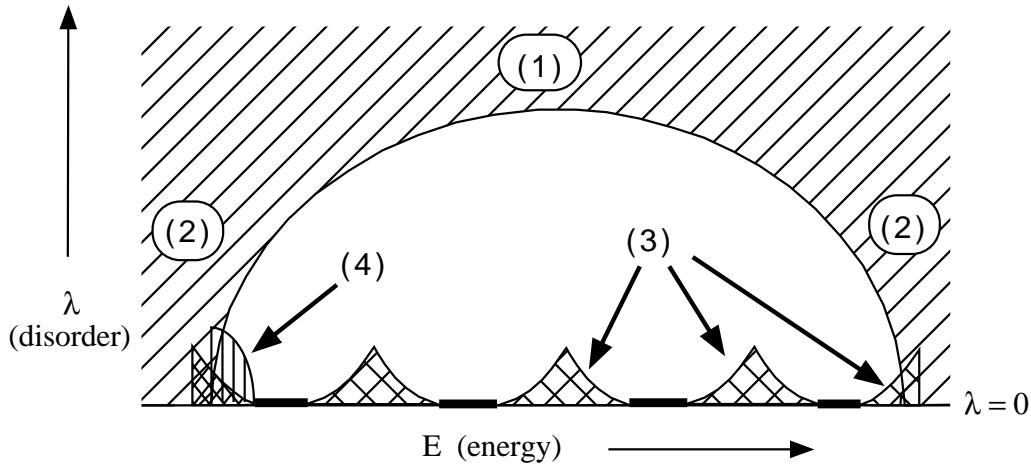  }
                        }
  {\footnotesize\caption{
  Different regions (schematically) in which localization
  occurs for an operator of the form $H=H_o + \lambda
  V_{random}$:  1) high disorder,
2) extreme energies, 3) weak disorder, away from
the spectrum of $H_o$, and 4) band edges.  The fractional moment
methods have been developed for the first three regimes.}}
  \label{fig:1}
 \end{center}
  \end{figure}

The condition (\ref{eq:gs}) was established for a broad class of systems,
in any dimension, under any of the three conditions: 1) high disorder,
2) extreme energies \cite{AizMo}, and 3) weak disorder \cite{Aiz94} away from
the spectrum of the unperturbed operator ($\lambda = 0$),
see Figure~\ref{fig:1}.
Localization is known also to occur at the band edges (case 4),
for which it can be proven \cite{Spencer,FigKl94a} by the
multiscale approach of  Fr\"ohlich and Spencer \cite{FS}.
However,  in this more delicate situation  condition \eq{eq:gs},
which leads to the implications
discussed below,  has not been established yet.
Neither has the condition been derived in the continuum
(for which localization results can be found in
refs.~\cite{HoMa,CH94,CH96,Wg,FiKl,Klopp,Kirsch,BdMG}).

We shall recapitulate below why \eq{eq:gs} can be viewed as a
natural technical expression of localization,
and present a heuristic derivation along
the lines of \cite{AizMo}. First, however let us list some of its
implications.

\masubsect{Physical implications of the resolvent condition}
\nopagebreak
There is a growing list of readily identifiable physical properties of
an electron gas which follow from \eq{eq:gs}, some of which have not
been derived without it.

A new statement which is added here to that list
is the exponential decay of the two--point function in the ground
state $|0\rangle\rangle $ if the Fermi energy,
$E_F$, falls within a range of energies for which eq.~(\ref{eq:gs})
holds. In terms of Fock--space Fermionic operators
\begin{equation}
\mbox{\boldmath $E$}( |\langle\langle
0|\psi^\dagger(x) \psi(y) |0\rangle\rangle | ) \le C{\rm
e}^{-\mu|x-y|}\;, \label{eq:tpf}
\end{equation}
or, in terms of the
relevant one--particle spectral projection $P_{\le E_F}$, on the
energy range $(-\infty, E_F]$,
\begin{equation}
\mbox{\boldmath $E$}
(|\langle x| P_{\le E_F} |y\rangle| ) \le C{\rm e}^{-\mu|x-y|}\;.
\label{eq:P}
\end{equation}

\noindent\remark i) If $E_F$ falls in a band of extended
states the above kernel decays by only a power law.

ii) The derivation of the exponential decay, given below, does not
require  eq.~(\ref{eq:gs}) to hold for all the energies below $E_{F}$.
Thus, it applies also to the case in which
the Fermi level is in a localized regime above a number of bands of
extended states.

iii)  The decay presented in eq.~(\ref{eq:P}) has interesting
implications on electrical conductance, both in the absence and in
the presence of magnetic field (Hall conductance), which are
discussed below.

Before we turn to the derivation of the conditions
(\ref{eq:gs}) and (\ref{eq:P}), let us list some other  physically
meaningful implications of eq.~(\ref{eq:gs}), some derived
by other authors, which it may be useful to have listed together.
These include:
\begin{itemize}
\item {\em Pure--point spectrum:\/} the spectrum of
the operator $H$ in the interval $[a,b]$ almost surely consists only
of (non--degenerate) eigenvalues with exponentially localized
eigenfunctions.
\end{itemize}
\noindent The implication is through either the
dynamical localization expressed in eq.~(\ref{eq:exp})
or the Kotani argument \cite{Ko},
as further explained by Simon-Wolff \cite{SiWo}. This argument
yields the useful principle that those decay properties of the
resolvent which hold for almost all energies in some interval ---
in a sense which is not affected by randomizations which
refresh the site  potentials --- are typically manifested also
by all the eigenfunctions the operator may have in that interval.
(The property of interest here is the exponential decay.)

\begin{itemize}
\item {\em Dynamical localization:\/}
 assuming \eq{eq:gs}, wave packets of states with energy in
the range $[a,b]$ do not spread.  The key estimate is (\cite{Aiz94})
 \begin{equation}
\mbox{\boldmath $E$}(\sup_t |\langle x| P_{[a,b]} {\rm e}^{-{\rm i}
tH} |y\rangle|) \le C{\rm e}^{-\mu|x-y|}\ \ .
\label{eq:exp}
\end{equation}
\end{itemize}
\noindent
One may note that eq.~(\ref{eq:exp}) is a stronger statement than
the exponential
localization of eigenfunctions (it was not available before
eq.~(\ref{eq:gs})). Implicit in its derivation is an
extension which permits to replace $e^{-itH}$ by an
arbitrary bounded function $f(H)$. Expressed in terms of the
eigenfunctions, with energies $E_n$ the assertion is:
\be
    \E(\sum_{E_n \in [a,b]} |\psi_n(x)| |\psi_n(y)| )
    \le C\e^{-\mu|x-y|}\;.
\label{eq:exp1}
\ee

\begin{itemize}
\item {\em Absence of level repulsion:\/} Minami \cite{Min} proved
that the localization condition \eq{eq:gs} implies that the {\em local}
distribution of the energy levels in the interval $[a,b]$,
for the system in $[-L,L]^{d}$, converges
(as $L\to \infty$) to the Poisson law --- i.e.,
the energy levels appear to be independently distributed.
\end{itemize}

The decay presented in eq.~(\ref{eq:tpf}) has interesting
implications on conductivity, both in the absence and in the presence
of magnetic field.

We refer here to the conductivities as given
by linear--response calculations.
We shall not address here the interesting questions concerning
the validity of such approximations, and the role of edge
currents.
\begin{itemize}
\item
{\em Vanishing of the d.c. electrical conductivity in the absence of
magnetic field:\/}
We shall see below that, for any dimension,
\be
\mbox{condition (\ref{eq:gs})}
\ \Longrightarrow \ \sigma_{i,j}(E_F) \ =\ 0 \ \
 \mbox{ (for $E_F\in (a,b)$)},
\label{eq:co0}
\ee
where $\sigma$ is the d.c. electrical conductivity
of an electron gas with Fermi energy $E_F$,
at the zero temperature limit and at zero magnetic field,
based a linear response calculation (Kubo formula):
\begin{equation}
\sigma_{i,j}(E_F)
=\lim_{\eta\downarrow 0}{\eta^2\over\pi} \sum_{x\in\mbox{\boldmath
$Z$}^d}x_ix_j \mbox{\boldmath $E$}(|G(0,x;E_F+{\rm i}\eta)|^2)
  \;. \label{eq:co}
\end{equation}
\end{itemize}
Let us remark that this expression for the conductivity
follows from the more standard Kubo  formula (\cite{Th74, Kubo})
under the assumption of finite conductivity.  (For completeness,
we present the argument in  Appendix~\Alphref{app:KF}.\ref{app:KF1}.)
An earlier proof of the
vanishing of $\sigma$ (in this form) was provided in ref.~\cite{FS}
for the regime covered by the ``multiscale analysis''.

In the presence of magnetic field, one is interested in the Hall
conductance.  The linear response calculation (see
Appendix~\Alphref{app:KF}.\ref{app:KF2})
is facilitated by the condition
\begin{equation}
\mbox{\boldmath $E$}\Bigl(\sum_{x\in\mbox{\boldmath
$Z$}^d}|x|^2| \langle 0| P_{\le E} |x\rangle|^2 \Bigr) <\infty\;,
\label{eq:x2}
\end{equation}
which is implied by \eq{eq:P}. The Kubo formula for this situation
is
\begin{equation}
\sigma_{i,j}(E)=
{\rm i}\,{\rm Tr}(P_{\le E}[[x_i,P_{\le E}],[x_j,P_{\le E}]])\;.
\label{eq:ku}
\end{equation}

\begin{itemize}
\item {\em Integral Quantum Hall Effect (2D):\/}
Bellissard, van Elst, and Schulz-Baldes \cite{BES} (BES)
proved that if \eq{eq:x2}
holds for a two--dimensional system then
at the corresponding Fermi energy the Hall
conductance $\sigma_{1,2}(E_{F})$ is an integer
$\times e^2/h$ and it is constant throughout intervals,
like $[a,b]$ of \eq{eq:exp}, in which
the localization length is is uniformly bounded.
\end{itemize}
Results leading to this conclusion were also developed by
Avron, Seiler and Simon ~\cite{AS2} (AS$^{2}$).

To put this result in a clearer perspective, consider the
continuum case of a Landau Hamiltonian weakly perturbed by
a random potential:
\be
H\ = \  \frac{1}{2} \left( \frac{\iu}{\hbar}  \underline{\nabla} +
      e \frac{\underline{B}\wedge \underline{x}}{2}
      \right)^2  +
      \lambda U_{random}(x)
\ee
where $U_{random}(x)$ could be of the form
 $U_{random}(x) = \sum_{j} \eta_j V(x-x_j)$,
with  $\{x_j\}$ randomly distributed points and $\{ \eta_j\}$
independent random coefficients.
Had all the results which are proven for the lattice Hamiltonians
been true in this case, one could deduce that for small
$\lambda$ the Hall conductance, as a function of the Fermi energy
exhibits  several plateaux, increasing as
$\sigma_{Hall} = (e^2/h) n $.  It would indeed be of interest to
see an extension to the continuum of the  localization
analysis discussed here.


The argument of BES~\cite{BES} is based on a number of
sophisticated results of non--commutative geometry
(in particular, theory developed by A. Connes).
To make these results more transparent, we
include below a direct and simple derivation
of the implication of \eq{eq:P} for the IQHE.
The discussion incorporates ideas which were developed by
AS$^{2}$~\cite{AS2}, discussed here in the context of
operators with random potentials
under the localization condition \eq{eq:P}.

\masect{Localization Bounds for the Resolvent}

To convey the key arguments leading to localization
bounds let us review the derivation of equation
\eq{eq:gs} for the situation with high disorder, or
at extreme energy (cases 1) and 2) in Fig.~\ref{fig:1}). Furthermore,
let us consider  the case where the magnetic field $B$ is
either zero or constant, $K_{x,y}$ is restricted to
the nearest--neighbor pairs (e.g., $K_{x,y} = $ the incidence
matrix), and $U^{per}_x \equiv 0$.
The equations
defining the resolvent (with $y=0$ and ${\rm i}\eta$
absorbed in $E$) is:
$ (H-E) G(x,0;E) = \delta_{x,0}$, or:
\be
(E-\lambda V_x)G(x,0;E) \ =\  \sum_{n\in\mbox{\boldmath $Z$}^d,|n|=1}
 K_{x,x+n} \; G(x+n,0;E) \;-\;
\delta_{x,0}  \; .
\label{eq:resolveq}
\ee
The solution of
this equation does not propagate well (it attenuates, or decays
exponentially) in regions where $E-\lambda V_x$ falls out of the
spectral range of the hopping operator seen on the right side,
i.e., where
$| E-\lambda V_x| > 2d$.  If $\lambda$ is large enough (or $E$ is very
large), most of the lattice will belong to this attenuation set, and
one may expect $G(x,0;E)$ to decay exponentially in $|x|$, indicating
exponential localization.

The big gap in this intuitive argument is that one still needs to
address the possible tunneling between the sparse sites at which $|
E-\lambda V_x| \le 2d$. This gap was closed by the ``multiscale
analysis'' of ref.~\cite{FS}, which was developed to handle the
technical problems caused by possible resonances, manifested through
small denominators. An alternative approach is to look at suitable
moments of $G(x,0;E)$, with the hope that the averaged quantities
will be informative enough. Here the small denominator problem shows up
as follows: for $H$ taken to be a finite matrix
(and $E$ real),
\begin{equation} {\rm Av}(|G(x,0;E)|)=\infty \label{blowup}
\end{equation}
where the average is either over the energy $E$,
integrated over any interval which intersects the spectrum,
or over the values of $V_x$.  The latter singularity is explained by
``rank--one perturbation formulae'' (or, alternatively, Cramer's rule),
e.g.,
\begin{equation} G(x,x;E)= \frac{1}{\widehat{G}(x,x;E)^{-1}+\lambda V_x}
\label{r1}
\end{equation}
where $\widehat{G}(x,x;E)$ --- the value of $G(x,x;E)$
for $V_x$  changed to $0$ ---  does not depend on $V_x$. In a
system in which $V_x$ is independent from the
values of the potential at other sites, there will be rare resonant
situations, in which the denominator in eq.~(\ref{r1}) is very small.
Despite its rarity, this phenomenon leads to ``$1/t$ tail'' in the
distribution of $|G(x,x;E)|$, as well as of the other matrix
elements, and to the divergence of the mean values expressed by
eq.~(\ref{blowup}).

Once the nature of the singularity is understood, one may see how to
keep it from obscuring the picture.  The key observation is that this
singularity does not cause blowups in fractional moments, i.e.,
averages of  $|G(x,0;E)|^s$ with any $0<s<1$. Averaging both sides of
equation (\ref{eq:resolveq}) raised to such a power $0<s<1$, one may
obtain the following relation (with the help of a decoupling
argument, discussed in \cite{AizMo,Aiz94})
\begin{equation}
c\  \mbox{\boldmath $E$}(|\lambda V_x -E|^s) \
\mbox{\boldmath $E$}(|G(x,0;E)|^s) \le
 \sum_{n\in\mbox{\boldmath $Z$}^d, |n|=1}
\mbox{\boldmath
$E$}(|G(x+n,0;E)|^s) \ +\  \delta_{x,0}  \; ,
\label{eq:averagedeq}
\end{equation}
with $c$ a finite constant depending on the
distribution of the random potential. In the above relation the
effect of the rare resonances is averaged out, and the simple
argument indicated below eq.~(\ref{eq:resolveq}) can be followed in a
conclusive way. When
\begin{equation} \gamma = \frac{2d}{c \;
\mbox{\boldmath $E$}(|\lambda V_x -E|^s)} < 1
\end{equation}
eq.~(\ref{eq:averagedeq}) implies that $\mbox{\boldmath
$E$}(|G(x+n,0;E)|^s)$ is a strictly subharmonic function on the
lattice (its value at $x$ is less that $\gamma (<1)$ times its
average over neighbors).  This readily implies the exponential decay
expressed in (\ref{eq:gs}), for strong disorder ($\lambda$ large), and
at extreme energies.  This argument leads to the following result.

\begin{thm} For a random Hamiltonian as in \eq{eq:h}, with
$V_{x}$ independent variables having the probability distribution
$r(v)dv$ with $r(v)$ bounded, if for some $0<s<1$ and $\mu>0$
\begin{equation}
\lambda>2\|r\|_\infty
\Big({2\over 1-s}\sum_{x\in\mbox{\boldmath $Z$}^d}
|K_{0,x}|^s{\rm e}^{\mu|x|}\Big)^{1/s}\;,
\end{equation}
then
\begin{equation}
\mbox{\boldmath $E$}( |G(x,y;z)|^s ) \le C{\rm e}^{-\mu |x-y|} \;,
\end{equation}
uniformly in $z\in\mbox{\boldmath $C$}\setminus\mbox{\boldmath $R$}$.
\label{thm:gs}
\end{thm}

The complete derivation is given below in Appendix~\Alphref{app:GF},
where we reproduce in a slightly streamlined fashion the argument of
ref.~\cite{AizMo}, and derive the exponential decay in a more general
setup, in which the probability distribution is not required to have a
density with respect to $dv$.
This approach can also be applied to other
regimes, where the resolvent can be studied through other equations,
e.g., the relation to the unperturbed resolvent operator:
$G=G_o - G_o \lambda V G$ (\cite{Aiz94}), yielding bounds which are
uniform in the two natural cutoffs: finite volume, and imaginary
energy shift {\rm i}$\eta$ (\cite{Gf}).

\masect{Exponential Decay for the Two--Point Function}
\label{sec:2pt}

We now turn to the implication of eq.~(\ref{eq:gs}) for the
two--point function, which for temperature $T\ge0$ is given by
\begin{equation}
\langle\psi^\dagger(x)
\psi(y)\rangle_T =\mbox{\boldmath $E$}(\langle
x|\theta_T(H-E_F)|y\rangle)\;,
\end{equation}
with the Fermi distribution
$\theta_T(u)=(1+\exp(u/T))^{-1}$.
For $T=0$: $\theta_0(u)=I[u\le 0]$.

At $T>0$ this function always exhibits exponential
decay, but at a rate for which the general bound --- independent of
$V$ and $E_F$ --- vanishes with $T$:
\begin{equation}| \langle
x|\theta_T(H-E_F)|y\rangle |\le C{\rm e}^{-\gamma T|x-y|}\;,
\end{equation}
(as follows from Theorem~\ref{thm:FH} below). We can say more under
the localization criterion \eq{eq:gs}:

\begin{thm}  If eq.~(\ref{eq:gs})
holds at the
Fermi energy $E_F$ then the two--point function decays exponentially
fast at the ground state (filled `Fermi sea'), and also at positive
temperatures --- with the correlation length ($\xi$) staying
uniformly bounded as $T\to 0$, i.e.
\begin{equation}
\mbox{\boldmath $E$}(|\langle
x|\theta_T(H-E_F)|y\rangle|) \le C{\rm e}^{-|x-y|/\xi}
\end{equation}
for all $T>0$, and also
\begin{equation} \mbox{\boldmath $E$}(
|\langle x| P_{\le E_F} |y\rangle| ) \le C{\rm e}^{-|x-y|/\xi} \;
\label{p_decay}
\end{equation}
(corresponding to the case $T=0$).
\label{thm:theta}
\end{thm}

\noindent {\bf Proof:}
To extract the information from the resolvents, it is convenient to employ
the contour integral
representation.  For the projection $P_{\le E_F}$ we chose
the path to consist of segments joining
$-\infty-{\rm i},\, E_F-{\rm i},\, E_F+{\rm i},\, -\infty+{\rm i}$.
Splitting the integral, we get
\be
P_{\le E_F}=Q_1(H-E_F)+Q_2(H-E_F) \; ,
\label{eq:cont}
\ee
\noindent with
\bea
Q_1(z)&=&{1\over 2\pi}\int_{-1}^1 d\eta\,{1 \over {\rm i}\eta-z}\;, \\
Q_2(z)&=&{1\over 2\pi{\rm i}}\int_{-\infty}^0 du
\Bigl[{1 \over u-{\rm i}-z}-{1 \over u+{\rm i}-z}\Bigr]\;.
\label{eq:Q}
\eea
(due to the randomness, the probability of there
being an eigenvalue exactly at the energy $E_F$ is zero (\cite{SiWo}),
a fact immediate from (\ref{eq:gs})).

For $T>0$, $\theta_T(z)$ has poles at
${\rm i}\pi T \times$\{odd integers\}, which grow
dense on the imaginary axis as $T\to 0$, with residues equal $(-T)$.
Evaluating the Cauchy integral along
the boundary $\Gamma$
of the strip $|{\rm Im}\,z|\le\eta$ with $\eta=
[\frac{1}{2\pi T}]\;2\pi T \approx 1$),
where $[X]$ denotes the integral part of $X$, one finds
$\theta_T(H-E_F)=\theta_{T,1}(H-E_F)+\theta_{T,2}(H-E_F)$ with
\bea
\theta_{T,1}(z)&=&{T}\sum_{n {\em odd}; |n\pi T|<\eta}
                    {1\over {\rm i} n\pi T - z}\;,\\
\theta_{T,2}(z)&=&{1\over 2\pi{\rm i}}\int_\Gamma
dw\,\theta_T(w) {1\over w-z}\nonumber\\
&=&{1\over 2\pi
{\rm i}}\int_{-\infty}^\infty du\,\theta_T(u)
\Bigl[{1\over u-{\rm i}\eta-z}-{1\over u+{\rm i}\eta-z}\Bigr]\;.
\label{eq:th2}
\eea
The sum defining $\theta_{T,1}(H-E_F)$ looks, at small
$T$, like a discrete approximation of the integral seen in
$Q_1(H-E_F)$.

In each case ($T=0$ and $T>0$), one may expect the
first term to be the more delicate one,
since it involves resolvents at arbitrarily small distances
(of the complex energies) from the real axis. Indeed, it is
at that point that the assumed condition eq.~(\ref{eq:gs}) enters.
For the more regular terms, $Q_2(H-E_F)$ and $\theta_{T,2}(H-E_F)$,
a starting point is provided by the Combes--Thomas estimate
(reproduced below in eq.~(\ref{eq:ct})), however we need to improve
on that in order to address the
question of the convergence of the resulting integrals.
(This question can be avoided in case the potential
is uniformly bounded ($H \ge E_0 > -\infty$) by
closing the contour at any point below $E_0$.)

More explicitly, we estimate the first term as follows:
\bea
     \E(|\langle x| Q_1(H- E_F)|y\rangle|)&
  \le& {1\over 2\pi}\int_{-1}^1 d\eta
  \E( |G(x,y;E_F+{\rm i}\eta)| )\nonumber
\\
&&\hskip -2cm\le {1\over 2\pi}\int_{-1}^1 d\eta\,\eta^{-(1-s)}
\E(|G(x,y;E_F+{\rm i}\eta)|^s )
\le C\e^{-\mu |x-y|}\;,
\eea
where we combined the assumed resolvent condition eq.~(\ref{eq:gs}),
with the general operator bound $|G(x,y;E_F+{\rm i}\eta)|\le\eta^{-1}$.
A similar estimate holds for
$\mbox{\boldmath $E$}(|\langle x|\theta_{T,1}(H)|y\rangle|)$,
the difference being that the integral over $\eta$ is replaced
by a corresponding  Riemann sum.

The exponential decay for the corresponding second pair of terms,
$\mbox{\boldmath $E$}(|\langle x| Q_2(H-E_F)|y\rangle|)$
and  $\mbox{\boldmath $E$}(|\langle x| \theta_{T,2}(H-E_F)|y\rangle|)$,
is a direct consequence of the following general result,
whose proof is given here in Appendix~\Alphref{app:CT}.

\begin{thm}  Let $H$ be as in eq.~(\ref{eq:h}) and $F$ be a
function analytic and bounded in the strip
$|{\rm Im}\,z|<\eta$ (by $ \|F\|_\infty$).
Then
\begin{equation}
|\langle x|F(H)|y\rangle|\le 18 \sqrt{2} \|F\|_\infty \ {\rm e}^{-\mu|x-y|}
\end{equation}
for any $\mu$ such that the quantity
$b(\mu)=\sum_{x\in\mbox{\boldmath $Z$}^d}|K_{0,x}|$ $({\rm e}^{\mu|x|}-1)$
satisfies $b(\mu) \le \eta/2$.
\label{thm:FH}
\end{thm}

In fact, this result relies on $F$ having a representation (\ref{eq:ct3})
analogous to (\ref{eq:Q}, \ref{eq:th2}).
\rightline{$\Box$}

\masect{Hall -- Kubo conductance as a charge--transport index}
\label{sec:HK}

The analysis described above applies in particular to systems with a uniform
magnetic field.  This case has, of course, attracted  a great
deal of attention due to the remarkable phenomenology associated with the
Quantum Hall Effect (QHE).
The Integral QHE (\cite{KDP}) (unlike the Fractional case \cite{TSG})
is understood now to be accountable for within the electron-gas picture,
in which the particles (or excitations) are subject to a one-particle
effective Hamiltonian of the type considered here  (see, e.g. \cite{PrGi}).

It has been pointed out that under suitable circumstances the values of
$\sigma_{H} h /e^2$ express topological indices, which would account
for both the observed integer values and for the robustness of the
phenomenon of IQHE (\cite{TKN2,ASS83,NTW,AS85}).
Curiously, the robustness is re--expressed in the
fact that similar conclusions are reached through different explanations,
in which the topological aspect of Hall conductance appears in different
disguises.

In this section we recount one of the approaches to Hall
conductance in two dimensions, employing the charge transport
index which was introduced and used very effectively
by Avron, Seiler and Simon (AS$^{2}$) \cite{AS2}.  The only addition
in this paper to the above work is the
derivation of the exponential decay of the kernel
of the projection operator $P_{\le E_F}$, \eq{p_decay}, which
(in a weaker form, e.g. fast enough power law) is essential for the
integrality of Hall conductance.  Another approach, which was developed
by Bellissard, van Elst, and Schulz-Baldes (BES) \cite{BES} will be
mentioned in the next section.  BES proceeded along a slightly different
path, employing the Chern--character view of the Hall conductance
and theorems proven in the context of ``non-commutative geometry''.
The latter work had a more intense focus on random
systems, and stated conditions under which Hall plateaux exist. However,
as noted in both works, the seemingly parallel tracks actually meet,
through a formula discovered by A. Connes.

The first step may be the formulation of a mathematical expression for
the Hall conductance within the model considered here. One intriguing
option is based on the charge pump mechanism proposed by Laughlin
\cite{Laugh81}.  Consider a system in
which the charges are confined to a plane (e.g., a suitable
interface), and the magnetic field is changed  through an adiabatic
process which results in the increase of the flux through a finite
region $D$ by $\Delta \Phi$.
Changes in the magnetic field are accompanied by an electric field
($\underline{E}$), including in the area surrounding $D$, and current ---
whose density we denote by $\underline{J}$.
The rate of charge transport across a contour
${\mathcal C}$ encircling $D$ is
\bea
\frac{\Delta Q}{\Delta t} \ & = & \ \oint_{\mathcal C}
 \underline{J} \cdot  \underline{n} \ d\ \ell \nonumber \\
 & = &  \  \sigma_D \oint_{\mathcal C}
 \underline{E} \cdot  \underline{n} \ d\ \ell \
 - \sigma_H \oint_{\mathcal C}
 \underline{E} \cdot  \underline{d\ \ell}
 \; ,
\label{ch.transp}
\eea
where $\sigma_D$ and $\sigma_H$ are elements of the bulk (homogenized)
conductivity tensor  (within the plane)
\be
\underline{J} = \sigma \underline{E} \; , \; \; \sigma\ = \
\left( \begin{array}{cc}
        \sigma_D & -\sigma_H  \\
        \sigma_H & \sigma_D
\end{array} \right) \; .
\ee
The last integral on the right side of \eq{ch.transp} (the induced
electromotive force) is tied, by Lenz's law, to the flux
change $-d \Phi / d t$.  The first term vanishes in situations in
which the direct conductivity ($\sigma_D$) is zero.   In that case,
the integral over time yields an expression for the Hall
conductance as the ratio of the transported  charge to the flux change:
\be
\sigma_H = \frac{\Delta Q}{\Delta\Phi} \;  .
\ee
One may note that it might be easier to analyze increments of flux
in multiples of $h/e$, since the
addition of such a flux quantum can be accomplished by means of a
gauge transformation, e.g.,
\be
U_a\ \psi (x) \ = \ e^{- \iu \theta_a(x)} \psi (x)
\ee
where $a\ (\notin \Z^2)$ is the location of the added flux line
and $\theta_a(x)$ is the angle of sight from $a$ to $x$ ($\Arg(x-a)$,
in the terminology of the complex plane).
The natural geometry for a charge pump based on this mechanism
is the Corbino disk, where
the transfer occurs between two conducting rings,
with the region between them filled by material
whose microscopic structure is modeled by the system discussed in
this paper.  Detailed analysis of this effect was presented
in the works of Laughlin \cite{Laugh81} and Halperin \cite{Halp82}.

Motivated by considerations related to the above discussion,
Avron, Seiler and Simon \cite{AS2} proposed an interesting
representation of the Hall conductance in the model discussed here,
in the $T=0$ limit.
They prove that {\em if the two--point function
$\langle x| P_{\le E_F} |y\rangle$ decays fast enough}, then
in a well defined sense only a finite number of states
are moved across the Fermi level.   The mathematical
expression of this is an index, which for a pair of projections
$P$ and $Q$ of compact difference
is defined as:
\bea
 \Ind(P,Q)  \ :=  \
 \dim \left\{\psi \in {\mathcal H}\ {\Huge |} \
\begin{array}{c}
        P\psi = \psi  \\
        Q\psi = 0
\end{array} \right\} \ -\
 \dim \left\{\psi \in {\mathcal H}\   {\Huge |}  \ \begin{array}{c}
        P\psi = 0  \\
        Q\psi = \psi

\end{array}   \right\} \; ,
\eea
(if $(P-Q)$ is a compact operator, the above
dimensions are finite).

Assuming that the above charge transport index
coincides also with the charge transferred
in the course of an adiabatic transition from $H$ to $UHU^*$,
($\Delta\Phi=h/e$),
AS$^2$ take for the Hall conductance the quantity
\be
\sigma_H \ = \ \frac{e}{h /e} \
   \E\left( \mbox{Index}(P_{\le E_F},
   U_a P_{\le E_F} U_a^*) \right) \; .
\label{hall=index}
\ee
The AS$^2$ study of this quantity rests on the following gem which
they added to the theory of Hilbert space operators.

\begin{thm} (\cite{AS2})
Let $P$ and $Q$ be a pair of orthogonal projections in a separable
Hilbert space ${\mathcal H}$, whose difference $P-Q$ is a compact operator.
If for some integer $n\ge 0$ the operator $(P-Q)^{2n+1}$ is trace
class, then (with no further dependence on $n$)
\be
{\rm tr\,} (P-Q)^{2n+1} \  =  \ \Ind(P,Q) \; .
\ee
\label{thm:index}
\end{thm}

This fact has a simple explanation through the observation that
the spectrum of $P-Q$ (which consists of  a collection of proper
eigenvalues in the interval $[-1, 1]$) is symmetric under sign
change ---
except for possible eigenvalues at $\pm 1$.  A particularly elegant
proof can be found in \cite{Av95}.

Further properties of the index are:
\begin{itemize}
\item[i)] additivity -
\be
\mbox{Index}(P,Q) \ + \ \mbox{Index}(Q,R) \ =
 \ \mbox{Index}(P,R) \; ,
\label{additivity}
\ee
for projections $P, Q, R$ which differ by compact operators,
and
\item[ii)] stability -
\be
\mbox{Index}(P,Q) \ = \ \mbox{Index}(P,UQU^{*}) \; ,
\ee
under unitaries $U$  with compact difference $(U - I)$.
\end{itemize}

\noindent
(AS$^2$ prove the above statements by reformulating
$\mbox{Index}(P,Q)$ as the Fredholm index of $PUP$ in
Range $P$, and invoking known properties of the latter.)

Two unitary operators $U_a$ and $U_b$, which
differ only at the location of the extra flux line,
are equivalent as far as the Hall
conductance \eq{hall=index} is concerned, since:
\begin{itemize}
\item[i)]
$(U_a U_b^{-1} - I)$ is a compact
operator and, by implication,
\item[ii)]
\be
\mbox{Index}(U_a P U_a^*, U_b P U_b^*) \ = \ 0 .
\label{index0}
\ee
\end{itemize}
\noindent
It follows that the charge transport index does not depend
(neither in its existence nor in its
value) on the location of the extra flux line.

The statement that some power of $T\ =\ P-U_a P U_a^*$ is trace class
may be verified by making use of:
\begin{lem}  For an operator with the matrix elements $T_{x,y}$
\be
\|T\|_3\equiv({\rm tr\,}|T|^3)^{1/3}\le
\sum_{b\in\Z^2}\left(\sum_{x\in\Z^2}|T_{x+b,x}|^3\right)^{1/3}\;.
\ee
\end{lem}
\noindent {\bf Proof:}
One may apply the norm's triangle inequality to the
decomposition $T=$\linebreak $\sum_{b\in\Z^2}T^{(b)}$, where
$T_{x,y}^{(b)}=T_{x,y}\delta_{x-b,y}$.  For each operator
in this sum:
$\|T^{(b)}\|_3=$\linebreak $\|T^{(b)*}T^{(b)}\|_{3/2}^{1/2}$ where
$(T^{(b)*}T^{(b)})_{x,y}=|T_{x+b,x}|^2\delta_{x,y}$ is a diagonal
operator for which the norm calculation is elementary.
\hfill$\Box$

The lemma implies
\be
\E(\|T\|_3)\le
\sum_{b\in\Z^2}\left(\sum_{x\in\Z^2}\E(|T_{x+b,x}|^3)\right)^{1/3}\;.
\label{eq:trcl}
\ee
If the Fermi energy is at a value for which the localization bound
\eq{p_decay} applies,
then the condition $\E(\|T\|_3)\ < \infty$ is satisfied for
$T_{x,y}=P_{x,y}(1-e^{-i(\theta_{a}(x) - \theta_{a}(y))})$,
as is easily seen from the bound
\be
\E(|T_{x,y}|^3)\ \le \ \E(|P_{x,y}|)\
|1-e^{-i(\theta_{a}(x) - \theta_{a}(y))}|^3
 \le A\ e^{-\mu |x-y|}\
\frac{C\ |x-y|^3}{1+|x-a|^3} \; .
\ee
In this situation, the combination of eq. \ref{eq:trcl},
Theorem \ref{thm:index} and some elementary algebra, imply that:
\begin{itemize}
\item[i)] the charge-transport index is well defined,
\item[ii)] it is given by
\bea
\mbox{Index}(P_{\le E_F}, U_a P_{\le E_F} U_a^*) \ =
 \ {\rm tr\,} (P_{\le E_F} - U_a P_{\le E_F} U_a^*)^3   \hspace{1.5in}
 \label{eq:angles1} \\
  =   \ 2\iu\ \sum_{x,u,v\in \Z^2} P_{x,u}  P_{u,v}  P_{v,x}
 \left[\sin(\angle (u,a,x) ) \ + \ \sin(\angle (v,a,u) ) \ + \
 \sin(\angle (x,a,v)) \right]  \nonumber
\eea
with $a$ an arbitrary point in $\R^2 \backslash \Z^2$,
\item[iii)] the above takes an integer value (which does not depend
on $a$).
\end{itemize}

Furthermore, as noted (in slightly different contexts)
by Connes \cite{Connes}, BES and AS$^2$,
$\mbox{Index}(P_{\le E_F}, U_a P_{\le E_F} U_a^*)$
is a translation invariant function of the randomness (a consequence
of equations (\ref{index0}) and (\ref{additivity}) ).
Since this function is also measurable
and integrable, Birkhoff's ergodic theorem implies that
the index does not fluctuate, in the sense
that for almost every realization of the random potential it takes
the value given by its mean.  A significant corollary is  that even
the mean takes an integer value, i.e. the Hall conductance,
as represented by \eq{hall=index}, takes the values
\be
\sigma_H \ = \ \frac{e^2}{h} \ n \; \; , \; \; \mbox{with}\  n\in \Z \,
\ee
and is given by the formula
\be
\sigma_H \  =  \    \frac{e^2}{h} \ 2\iu
 \sum_{x,u,v\in \Z^2} \E \left( P_{x,u}  P_{u,v}  P_{v,x}\right)
 \left(\sin\alpha  \ + \ \sin\beta  \ + \
 \sin\gamma \right)  \; ,
\label{eq:angles2}
\ee
or, using translation invariance,
\be
\sigma_H \  =  \    \frac{e^2}{h} \ 2\iu
 \sum_{u,v\in \Z^2, a\in \Z^{2*}} \E \left( P_{0,u}  P_{u,v}
  P_{v,0}\right)
 \left(\sin\alpha \ + \ \sin\beta \ + \
 \sin\gamma \right)  \; ,
\label{eq:angles3}
\ee
where $\{ \alpha, \beta, \gamma\}$ are the angles described
explicitly in \eq{eq:angles1}, and in the second expression
the summation over $x$ is replaced by a sum over $a$, varying over the
dual lattice.

The above discussion leads also to the statement, formulated by
Bellissard et al. \cite{BES},
that the Hall conductance is constant in regions in which a
localization estimate, like our \eq{eq:P}, holds uniformly in $E$.
To see this result, it is convenient to first relate the above
expression for $\sigma_H$ with the other expression which was
proposed for it, in what is known as the Streda formula \cite{streda}.

\masect{Relation with the Streda-Kubo formula}
\label{sec:SK}

In his work on non-commutative geometry, A. Connes \cite{Connes}
presented a remarkable formula, whose discrete version
reads as follows.  For any $u,v \in \Z^2$:
\bea
\sum_{a\in \Z^{2*}}[\sin(\angle(u,a,0) )\ + \ \sin(\angle(v,a,u) )
\ + \ \sin(\angle(0,a,v) ) ] \ & =&  \pi \ (u_2 v_1 - u_1v_2)
\nonumber \\
 & = & \pi \  u \wedge v \; .
\label{eq:connes}
\eea
(For completeness, a streamlined derivation is included here in
Appendix~\Alphref{app:CF}.)

Using Connes' formula, the expression (\ref{eq:angles3})
derived for the Hall conductance starting from the charge
transport index is transformed to (with $P\equiv P_{\le E}$):
\bea
\sigma_H \ &  =  & \  \frac{e^2}{h} \ 2 \pi i
 \sum_{u,v\in \Z^2} \E \left( P_{0,u}  P_{u,v}  P_{v,0}\right)
 \ (u \wedge v) \label{sigma2} \\
 & = &  \    \frac{e^2}{h} \ 2\pi i \
   \E \left(
 \langle 0|P X_2 P X_1 P  |0 \rangle
 -  \langle 0|P X_1 P X_2 P  |0 \rangle
 \right) \nonumber \\
 &= & \  \frac{e^2}{h} \ 2\pi i \
   \E \left( \langle 0|P [ [X_2,P], [X_1,P] ] P|0 \rangle
   \right) \   \equiv  \ \sigma_{2,1}(E) \; .
\label{sigma3}
\eea
(By translation invariance, the rightmost projection in
\eq{sigma3} can be omitted.)

The above expressions are of interest for a number of reasons:
\begin{itemize}
\item[i)] In a suitable setup,
the expression provided by \eq{sigma3}
takes the form of a Chern number, and thus it offers another
perspective on the topological aspect of Hall conductance.
(This topic  will not be covered here, as it
is extensively discussed in references \cite{ASS83,Th94,BES}.)
\item[ii)]  The above expression coincides with the
Kubo formula \eq{eq:ku} for conductance, based on a linear response
calculation, e.g., like the one presented here in Appendix~\Alphref{app:KF}).
\item[iii)]  The  expression provided by \eq{sigma2} is very convenient
for the derivation of sufficient conditions for the continuity of the
Hall conductance, i.e., for the existence of plateaux.  The following
result is fashioned on a theorem formulated by Bellissard et al.
\cite{BES} (where the assumption is slightly different).
\end{itemize}

\begin{thm} {\em(Slightly modified version of a result in \cite{BES})}
For a random Schr\"odinger operator, $H = K_{x,y} + \lambda V_x$, with
$K$ incorporating a uniform magnetic field, as in \eq{eq:hb},
and $V_x$ a random potential whose probability distribution is
invariant and ergodic under translations,  $\sigma_H(E)$
(the zero-temperature Hall Conductance at Fermi-energy $E$)
is a constant integral multiple of $e^2/h$ throughout each interval
of energies $E$, over which for some $q>2$ the quantity
\be
\xi_q \ = \  \sum_{x \in \Z^2}\E\left(|\langle 0| \ P_{\le E}
 \ |x \rangle |^q \right)^{1/q} \ |x|  \;
\label{eq:length}
\ee
is uniformly bounded.
\end{thm}

This statement depends on the continuity of the integrated density of
states, $\E(\langle 0| \ P_{\le E} \ |0 \rangle) $,
a fact which is known for all translation invariant random
operators in the setting \eq{eq:h}, \cite{DelSou}.

\noindent{\bf Proof: }    By an elementary telescopic decomposition,
and an application of translation-invariance, \eq{sigma2} implies,
for $1/q + 1/q + 1/r = 1$ $(r\ge 1)$,
\bea
|\ \sigma_H(E+\Delta E) &-& \sigma_H(E)\ |   \  \le    \\
  & \le & 3  \frac{2 e^2}{h}
\sum_{u,v\in Z^2} \E\left( |P_{0,u}^{\#}|^q
\right)^{1/q}  \E\left( |P_{u,v}^{\#}|^q \right)^{1/q}
 \E\left( |\Delta P_{v,0}|^r\right)^{1/r} \ |u| |v-u|
\nonumber \\
& \le & \ \frac{6 e^2}{h} \
\left[ \sum_{u\in Z^2} \E\left( |P_{0,u}^{\#}|^q \right)^{1/q} \ |u|
\right]^2 \  \E\left( |\Delta P_{0,0}|\right)^{1/r}
\eea
where $P^{\#}$ is either $P_{\le E}$ or $P_{\le E+\Delta E}$,
$\Delta P = P_{\le E+\Delta E}-P_{\le E}=P_{(E,E+\Delta E]}$,
and use was made of the Cauchy--Schwarz inequality
$|\Delta P_{v,0}| \le |\Delta P_{0,0}| $.
The last factor on the right side tends to zero by the
afore-mentioned general
continuity results, while the other factors stay
bounded under the assumption \eq{eq:length} that the localization
lengths stays uniformly bounded.
\hfill$\Box$

An explicit estimate showing the continuity of the integrated
density of states
(though in less than full generality)
is the Wegner bound \cite{wegner}:
\be
 \E\left( |\Delta P_{0,0}|\right) \ \le
 \ \lambda^{-1}|\Delta E| \ \| r \|_{\infty} \; ,
\label{eq:wegner}
\ee
which is valid for random Hamiltonians where the probability
distribution for the potential has a bounded density function
$r(V) \ \le \| r \|_{\infty}$.

\startappendix
\maappendix{The Kubo Formula for the Electric Conductivity}
\label{app:KF}

In this Appendix we present the linear response calculations
leading to the expressions we invoked for the conductance.
In particular, we shall reconcile the expression (\ref{eq:co})
for $\sigma$ with another familiar form of the ``Kubo formula''.
We shall not address here the question of the validity of the linear
response theory, which requires a more thorough analysis.

To derive the Kubo linear response formula for conductivity in
a system of non-interacting Fermi particles consider switching an
electric field $\underline{ E}$ adiabatically  through
the time--dependent Hamiltonian
\begin{equation}
H(t)=H+\underline{E} \cdot \underline{x} {\rm e^{\eta t}}\;,
\qquad(-\infty<t\le 0,\,\eta\downarrow 0)\;.
\end{equation}
The unperturbed density matrix $\rho$ shall be in equilibrium
w.r.t. $H$, i.e.,
$[\rho,H]=0$. A typical example is the Fermi distribution
\begin{equation}
\rho=\theta_T(H-E_F)\;,\qquad (T\ge0)\;.
\label{eq:fd}
\end{equation}
The perturbed density matrix $\rho(t)$
satisfies the initial value problem
\begin{equation}
{d\over dt}\rho(t)=-{\rm i}[H(t),\rho(t)]\;,\qquad
\lim_{t\to-\infty}{\rm e}^{{\rm i}Ht}\rho(t){\rm e}^{-{\rm i}Ht}=\rho\;.
\label{eq:evolution}
\end{equation}

To first order in $\underline{ E}$ (the linear response theory)
the solution to \eq{eq:evolution} is
\begin{equation}
\rho(0)-\rho=-{\rm i}
\int_{-\infty}^0 dt\,{\rm e}^{\eta t}
{\rm e}^{{\rm i}Ht}[{\underline{ E}}\cdot \underline{ x},
\rho]{\rm e}^{-{\rm i}Ht}\;.
\end{equation}
For the current density due to the field this yields
\begin{equation}
\underline{j}=-{\rm Tr}(\underline{v}(\rho(0) - \rho) )\;,
\end{equation}
where $\underline{v}={\rm i}[H,\underline{x}]$
is the velocity and ${\rm Tr}$ denotes the trace per
unit volume:
\begin{equation}
{\rm Tr\,}A=\lim_{\Lambda\uparrow\mbox{\boldmath $Z$}^d}
|\Lambda|^{-1}\sum_{x\in\Lambda}\langle x|A|x\rangle
\;.
\label{eq:tr}
\end{equation}
Therefore  $j_i=\sigma_{i,j}{ E}_j$, with
\begin{equation}
\sigma_{i,j}=-\lim_{\eta\downarrow 0}\,{\rm Tr}(
\int_{-\infty}^0 dt\,{\rm e}^{\eta t}{\rm e}^{-{\rm i}Ht}[H,x_i]
{\rm e}^{{\rm i}Ht}[x_j,\rho])\;.
\label{eq:a2}
\end{equation}
Translation invariance permits one to replace here ${\rm Tr}$ by
an average over the disorder of the diagonal term:
\begin{equation}
{\rm Tr\,}A=\mbox{\boldmath $E$}(\langle 0|A_\omega|0\rangle)\;.
\label{eq:trace}
\end{equation}

The ergodic argument enabling \eq{eq:trace} was presented
in similar context in ref.~\cite{BES}.
Let us consider the probability space whose points are the
random environments, i.e. potentials
$\omega=\{V_x\}_{x\in\mbox{\boldmath $Z$}^d}$,
and let $T_a\omega=\{V_{x-a}\}_{x\in\mbox{\boldmath $Z$}^d}$ be the
shift by
$a\in\mbox{\boldmath $Z$}^d$. We note that $T_a$
act as ergodic shift.
An observable $A=\{A_\omega\}_{\omega\in\Omega}$
is stationary if
\begin{equation}
U(a)A_\omega U(a)^{-1}=A_{T_a\omega}\;,
\end{equation}
for all vectors $a$ which are periods of $U^{per}_x$. Here $U(a)$ are the
magnetic translations (\ref{eq:m1}). The Hamiltonian (\ref{eq:h}) is
stationary in
this sense. For stationary operators with
$\langle x|A_\omega|y\rangle=0$ for
$|x-y|$ large enough and all $\omega\in\Omega$
the trace (\ref{eq:tr}) almost surely takes the value given by
the right side of equation~(\ref{eq:trace}).
That expression is valid for $U^{per}=0$, otherwise
$|0\rangle$ should be replaced by an average over $|x\rangle$ with
$x$ ranging over a unit cell.
Note that this defines a linear, positive, commutative functional of $A$.
This functional then naturally defines a trace class of operators,
to which \eq{eq:trace} extends by continuity.

We now separate the discussion into two cases: zero magnetic field
where the system is time reversal invariant, and
non-vanishing magnetic field, in which case we are interested in the
Hall conductance.

\masubsectapp{Time reversal invariant systems}
\label{app:KF1}

At non-zero temperatures the density matrix is of the form $\rho=f(H)$
with a  smooth function $f$ with $\|f'\|_1<\infty$.  In such case,
for time reversal invariant systems, where (\ref{eq:a2}) is symmetric
in $\{i,\,j\}$,  \eq{eq:a2} can be brought to the form
\begin{equation}
\sigma_{i,j}=-\pi\lim_{\eta\downarrow 0}
{\rm Tr}\Big(\int\!\!\int \delta_\eta(\lambda-\mu)
{f(\lambda)-f(\mu)\over \lambda-\mu}dP_{\le\lambda}v_idP_{\le\mu}v_j\Big)\;,
\label{eq:a3}
\end{equation}
where $\delta_\eta(x)=(\eta/\pi)(x^2+\eta^2)^{-1}$.
Equation~(\ref{eq:a3}) corresponds to a well-familiar form of the Kubo
formula (\cite{Th74,Pas}).

We shall now see that this formula yields the expression seen in
eq.~(\ref{eq:co}):
\begin{equation}
\tilde{\sigma}_{i,j}(E)
=\lim_{\eta\downarrow 0}{\eta^2\over\pi} \sum_{x\in\mbox{\boldmath
$Z$}^d}x_ix_j \mbox{\boldmath $E$}(|G(0,x;E+{\rm i}\eta)|^2)
  \;. \label{eq:a4}
\end{equation}
(tilde added to avoid confusion).
The main assuption we shall use is that $\tilde{\sigma}_{i,j}(E)$
is finite {\em at all energies}. Rougly speaking,
this corresponds to a situation where the motion in absence of the
electric field is diffusive, at most.  On physical grounds one may
expect that to be the case in the presence of disorder,
regardless of localization.  Let us remark that a related expression
for the conductance can be based on the Einsten relation
of conductance with the diffusion constant  (\cite{Kubo}), however
we do not use this relation here.

\begin{thm}
Assume that
\begin{equation}
\sup_{\eta>0}\ \eta^2\sum_{x\in\mbox{\boldmath
$Z$}^d}x^2 \mbox{\boldmath $E$}(|G(0,x;E+{\rm i}\eta)|^2) \ \le
Const.
\label{eq:a5}
\end{equation}
with a finite constant which applies for all energies.
If  the limit in eq.~(\ref{eq:a4}) exists for all $E$, then
\begin{equation}
\sigma_{i,j}=-\int f'(E)\tilde{\sigma}_{i,j}(E)dE\;,
\label{eq:a6}
\end{equation}
where $f(E)=\theta_T(E-E_F)$ and
$\sigma_{i,j}=\sigma_{i,j}(T)$ ($T>0$).
For $T\to 0$, in lieu of the last assumption
we require that the limit
$\eta \to 0$ in eq.~(\ref{eq:a4}) exists for $E$ in some
neighbourhood of $E_F$, and is continuous there.
Under these assumptions,
\begin{equation}
\lim_{T\downarrow 0}\sigma_{i,j}\ =\ \tilde{\sigma}_{i,j}(E_F)\;.
\label{eq:a6a}
\end{equation}
\end{thm}

Clearly,  eq.~(\ref{eq:a6a}) is the limiting expression for
eq.~(\ref{eq:a6}) as $T\downarrow 0$, where $f$ becomes a step
function.  However a small clarification  may be needed,  since 
under the weaker assumption made for $T=0$ the 
limit in eq.~(\ref{eq:a3})  may exist only  for 
subsequences  $\eta_n \to 0$.  Nevertheless, equation~(\ref{eq:a6a})
means that the double limit $\lim_{T \to 0} \lim_{\eta \to 0} \sigma_{i,j}$ is 
unambiguous.

\noindent {\bf Proof:} We first address the r.h.s. of (\ref{eq:a6}):
$\tilde{\sigma}_{i,j}(E)$ is the limit as $\eta\downarrow 0$ of
\begin{eqnarray}
{\eta^2\over\pi}\sum_{x\in\mbox{\boldmath
$Z$}^d} \hskip -0.6cm&&
x_ix_j \mbox{\boldmath $E$}(|G(0,x;E+{\rm i}\eta)|^2)\nonumber\\
=&&-{\eta^2\over\pi}\E(\langle 0|
[x_i, (H-E+{\rm i}\eta)^{-1}]\,[x_j,(H-E-{\rm i}\eta)^{-1}]
|0\rangle) \nonumber\\
=&&{\eta^2\over\pi}{\rm Tr\,}
[(H-E+{\rm i}\eta)^{-1}v_i (H-E+{\rm i}\eta)^{-1}
(H-E-{\rm i}\eta)^{-1}v_j(H-E-{\rm i}\eta)^{-1}]\nonumber\\
=&&\pi\int\!\!\int \delta_\eta(\lambda-E)\delta_\eta(\mu-E)
m_{i,j}(d \lambda,d \mu)\;.
\end{eqnarray}
where $m_{i,j}(d \lambda,d \mu)={\rm Tr}(dP_{\le\lambda}v_idP_{\le\mu}v_j)$
is the conductivity measure (\cite{Pas}). Under the assumption
(\ref{eq:a5}), the limit $\eta\downarrow 0$ may be interchanged
with the $E$-integration seen on the right side of \eq{eq:a6}.
We thus obtain
\begin{equation}
\sigma_{i,j}+\int f'(E)\sigma_{i,j}(E)dE=
\lim_{\eta\downarrow 0}\,\pi
\int\!\!\int g_\eta(\lambda,\mu) m_{i,j}(d \lambda, d \mu)
\label{eq:a7}
\end{equation}
with
\begin{equation}
g_\eta (\lambda,\mu)=
\int dE\int_0^1 ds[f'(E)-f'(s\lambda+(1-s)\mu)]
\delta_\eta(\lambda-E)\delta_\eta(\mu-E)\;.
\label{eq:a8}
\end{equation}
The claim \eq{eq:a6} is thus equivalent to the assertion that
that the right side of \eq{eq:a7} vanishes.
Let us note that $m_{i,j}(d \lambda, d \mu)$ is a finite measure.
Using nothing more than the smoothness of $f$, one can show
(we skip the analysis here) that: i)
$g_\eta (\lambda,\mu)$ is uniformly bounded, and ii) it tends to
zero pointwise as $\eta\downarrow 0$. Thus, (\ref{eq:a7}) vanishes
by dominated convergence.

The zero-temperature limit follows by elementary analysis.
\hfill$\Box$

\masubsectapp{Systems with decaying two-point function}
\label{app:KF2}

Let us return now to the expression \eq{eq:a2} for conductance,
and discuss it in the presence of magnetic field.   Our goal is
to replace it by a more explicit formula.
We focus on the zero-temperature limit, where
$\rho=P_{\le E_F}$.  The following argument, which is related to
one presented in [20], applies under
the (localization) assumption of rapid decay of the matrix elements
$\langle 0| P_{\le E} |x\rangle $.

The Hilbert--Schmidt ideal
\begin{equation}
{\cal T}=\{A\mid{\rm Tr\,}A^*A<\infty\}
\end{equation}
is a Hilbert space with inner product
$(A,B)={\rm Tr\,}A^*B$. The linear map $\cal L_H$ on $\cal T$
given by ${\cal L}_H(A)=[H,A]$ is self--adjoint. Its resolvent
is seen to be
\begin{equation}
({\cal L}_H+{\rm i}\eta)^{-1}(A)=
-{\rm i}\int_{-\infty}^0 dt\,{\rm e}^{\eta t}
{\rm e}^{-{\rm i}Ht}A{\rm e}^{{\rm i}Ht}
\end{equation}
for $\eta>0$.
Thus, eq. (\ref{eq:a2}) has the appearance of:
\begin{equation}
\sigma_{i,j}=-{\rm i}\,\lim_{\eta\downarrow 0}\,{\rm Tr}(
({\cal L}_H+{\rm i}\eta)^{-1}{\cal L}_H(x_i)[x_j,\rho])\;.
\label{eq:a2*}
\end{equation}
In this form, one is tempted to apply the
spectral theorem, which implies that
\begin{equation}
\lim_{\eta\downarrow 0}({\cal L}_H+{\rm i}\eta)^{-1}{\cal L}_H(A)=A
\label{eq:spectral}
\end{equation}
for $A\in({\rm Ker\,}{\cal L}_H)^\perp$.
However $x_i$ is not even in the space $\cal T$, and hence neither
\eq{eq:a2*} nor \eq{eq:spectral} applies.
In the following argument
this difficulty is resolved through the replacement of $x_i$
by $[[x_i,P],P]$.

At zero temperature $\rho=P_{\le E}\equiv P$ is a projection and we
have
$[x_j,P]=P[x_j,P]$ $(1-P)+(1-P)[x_j,P]P$.  The substitution of this into
\eq{eq:a2} amounts, by cyclicity, to the substitution of  $x_i$
there by the following expression
\begin{equation}
(1-P)x_iP+Px_i(1-P)=[[x_i,P],P]\;.
\end{equation}
Unlike $x_i$, the above quantity is in the space $\cal T$
provided
\begin{equation}
{\rm Tr}([x,P]^*[x,P])=
\mbox{\boldmath $E$}\Bigl(\sum_{x\in\mbox{\boldmath
$Z$}^d}|x|^2| \langle 0| P_{\le E} |x\rangle|^2 \Bigr) \ < \ \infty
\; .
\end{equation}
Furthermore, in that case $[[x_i,P],P]$  is also in
$({\rm Ker\,}{\cal L}_H)^\perp$,
since for any $B\in\mbox{\rm Ker}{\cal L}_H$ we have
${\rm Tr}(B^*[[x_i,P],P])={\rm Tr}([P,B^*][x_i,P])=0$.
That makes \eq{eq:spectral} applicable, and the conclusion is
\begin{equation}
\sigma_{i,j}(E)=-{\rm i}\,{\rm Tr}([[x_i,P],P][x_j,P])
={\rm i}\,{\rm Tr}(P[[x_i,P],[x_j,P]])\;.
\end{equation}

Note that $\sigma_{i,j}(E)$ is antisymmetric and that, in particular,
the longitudinal conductivity vanishes. In $d=2$ it is an
integer divided by
$2\pi$ by results of \cite{BES, AS2} and reviewed here in Sections
\ref{sec:HK}, \ref{sec:SK}.
If the Hamiltonian (\ref{eq:h}) is time reversal invariant,
which requires the absence of a magnetic field, the tensor $\sigma_{i,j}(E)$
is also symmetric and hence vanishes altogether.

\maappendix{Exponential Decay for the Green Function} \label{app:GF}

Following is a rigorous derivation of eq.~(\ref{eq:gs}) for high
disorder, along the lines of ref.~\cite{AizMo} but with somewhat
more explicit bounds. As mentioned already, the argument can be
extended also to other regimes. We allow the probability
distribution to be of a more general type than considered in
Theorem~\ref{thm:gs}.

\medskip\noindent {\bf Definition }{\it Let $0<\tau\le 1$.
A $\tau$--regular
measure $q(dv)$ is one satisfying
\begin{displaymath}
  q[v-\delta,v+\delta]\le\ \mbox{\em Const.\/} \delta^\tau
\end{displaymath}
for all $v\in\mbox{\boldmath $R$},\,\delta>0$, in which case we let
$M_{\tau}(q)$ be the optimal (smallest) choice for the
constant {\em Const.} }

\medskip
Such a measure need not have a density $dq/dv\equiv r(v)$. If it
does, with $r\in L^p(\mbox{\boldmath $R$})$ for $p=(1-\tau)^{-1}$,
then
$M_{\tau}(q)\le 2^\tau\|r\|_p$.  Following is the
localization statement.

\medskip\noindent{\bf Theorem \ref{thm:gs}'\ }
{\it Let $0<s<\tau$ and $\mu>0$. If
\begin{equation}
\lambda>M_{\tau}(q)^{1/\tau}
\Big(C_{s,\tau}^{-1}\sum|K_{0,x}|^s{\rm e}^{\mu|x|}\Big)^{1/s}\;,
\label{eq:hy}
\end{equation}
where $C_{s,\tau}=(2\tau)^{-1}(\tau-s)$, then
\begin{equation}
\mbox{\boldmath $E$}( |G(x,y;z)|^s ) \le C{\rm e}^{-\mu |x-y|} \;,
\label{eq:gs1}
\end{equation}
uniformly in $z\in\mbox{\boldmath $C$}\setminus\mbox{\boldmath $R$}$.}

\medskip\pagebreak
We begin the proof by stating the following auxiliary fact
which plays the role of the decoupling lemmas of
ref.~\cite{AizMo,Aiz94}. Its proof is given in the next Appendix.

\begin{lem} Let $0<s<\tau$. Then
\begin{equation}
  \int dq(v){|v-\alpha|^s\over|v-\beta|^s}
     \ge C_{s,\tau}\Bigl({\int dq\over M_{\tau}(q)}\Bigr)^{s/\tau}
      \int dq(v){1\over|v-\beta|^s}
\label{eq:dl}
\end{equation}
for all $\tau$--regular measures $dq,\,0\not\equiv dq\geq 0$ and all
$\alpha,\,\beta\in\mbox{\boldmath $C$}$.
\label{lem:2}
\end{lem}

\medskip
Below we will also need a simple upper bound for the r.h.s. We split
$\mbox{\boldmath $R$}$ into
$|v-\beta|^{-s}\le\lambda$ and its complement. This gives
\begin{equation}
\int dq(v)|v-\beta|^{-s}
\le\lambda\int dq(v)
  +\int_\lambda^\infty d\lambda'q[|v-\beta|^{-s}\ge\lambda']
\le{\tau\over \tau-s}M_{\tau}(q)^{s/\tau}\Bigl(\int
dq(v)\Bigr)^{1-(s/\tau)}\;,
\label{eq:in}
\end{equation}
where we minimized over $\lambda>0$.

For the following argument it is important to know that the
resolvent is a simple rational function of each of the
the potential parameters ($V_x$) at fixed values of the others.
For matrices that is easily seen from Cramer's formula.
More generally, let
$P_x=|x\rangle\langle x|$ and let $\widehat{H}$ be the Hamiltonian
(\ref{eq:h}) for $V_x$ changed to $0$. From the resolvent identity
\begin{equation}
(\widehat{H}-z)^{-1}=(1+V_x (\widehat{H}-z)^{-1}P_x)(H-z)^{-1}
\label{eq:re}
\end{equation}
we get
\begin{equation}
G(x,y;z)={1\over V_x+\widehat{G}(x,x;z)^{-1}}
  \cdot{\widehat{G}(x,y;z)\over\widehat{G}(x,x;z)}\;.
\label{eq:rk}
\end{equation}
A simple application thereof is
\begin{equation}
\sup_{z\in\mbox{\boldmath $C$}\setminus\mbox{\boldmath $R$}}
\mbox{\boldmath $E$}(|G(x,x;z)|^s)<\infty\;.
\label{eq:apb}
\end{equation}
In fact already the expectation w.r.t. $V_x$ is uniformly bounded by
(\ref{eq:in}).

\medskip\noindent
{\bf Proof of Theorem~\ref{thm:gs}':}
With no loss of generality we set $y=0$ and consider
eq.~(\ref{eq:resolveq}) or, more precisely, its replacement
for the more general situation (\ref{eq:h}):
\begin{equation}
(z-\lambda V_x-U^{per}_x)G(x,0;z) =
\sum_{y\in\mbox{\boldmath $Z$}^d}
 K_{x,y} \; G(y,0;z) \;-\;\delta_{x,0}  \; .
\label{eq:resolveq1}
\end{equation}
To ensure existence we took the resolvent at energies
$z\in\mbox{\boldmath $C$}\setminus\mbox{\boldmath $R$}$.
Raising eq.~(\ref{eq:resolveq1}) to the power $0<s<1$ yields
\begin{equation}
|z-\lambda V_x-U^{per}_x|^s|G(x,0;z)|^s\le
\sum_{y\in\mbox{\boldmath $Z$}^d}
|K_{x,y}|^s|G(y,0;z)|^s\;,\qquad(x\neq 0)\;.
\label{eq:resolveqs}
\end{equation}
Note the particular dependence
(\ref{eq:rk}) of $G(x,0;z)$ on $V_x$.
Upon taking expectations and using
eq.~(\ref{eq:dl}) we obtain
\begin{equation}
a \mbox{\boldmath $E$}(|G(x,0;z)|^s)\le
\sum_{y\in\mbox{\boldmath $Z$}^d}
|K_{x,y}|^s\mbox{\boldmath $E$}(|G(y,0;z)|^s)\;,\qquad(x\neq 0)\;.
\label{eq:sh1}
\end{equation}
with $a=C_{s,\tau}M_{\tau}(q)^{-s/\tau}\lambda^s$.

When $a > \sum_{y\in\mbox{\boldmath $Z$}^d}|K_{0,y}|^s$,
the above is a subharmonicity statement for the function
$g(x)=\mbox{\boldmath $E$}(|G(x,0;z)|^s)$, which combined with
uniform boundedness and exponential decay of $|K_{x,y}|^s$
is known to lead to exponential decay.
Following is one of the many methods to reach that conclusion
(another can be found in ref.~\cite{AizMo}). It is based on subharmonic
comparison.

For a provisional uniform bound let us note that
$\|(H-z)^{-1}\|\le|{\rm Im}\,z|^{-1}$ yields:
\be
g(x) \le |{\rm Im}\,z|^{-s} \; .
\ee
Thus, $g(x)$ can be viewed as an element in the space of
bounded  functions $\ell^\infty(\mbox{\boldmath $Z$}^d)$,
and \eq{eq:sh1} can be recast as
\begin{equation}
a\ g(x)\le(hg)(x)\;,\qquad(x\neq 0)\;,
\end{equation}
where $h$ is the operator with the kernel $|K_{x,y}|^s$.
Note that if $\varphi\in\ell^\infty(\mbox{\boldmath $Z$}^d)$
with $\varphi(0)\le0$ satisfies
\begin{equation}
a\varphi(x)\le(h\varphi)(x)\;,\qquad(x\neq 0)
\end{equation}
with $a>C=\sup_{x\in {\boldmath Z}^d}
\sum_{y\in\mbox{\boldmath Z}^d}|K_{x,y}|^s$, then
$\varphi(x)\le0$. In fact, if
$M=\sup_{x\in\mbox{\boldmath $Z$}^d}\varphi(x)>0$ then
$a\varphi(x)\le CM$
and we get a contradiction by taking the supremum over $x$.
We apply this
conclusion to
$\varphi(x)=g(x)-g(0){\rm e}^{-\mu|x|}$.
The length scale $\mu^{-1}$ is set by the condition
\be
\sum_{y\in\mbox{\boldmath $Z$}^d}|K_{0,y}|^s{\rm e}^{\mu|y|}<a \; ,
\ee
which is eq. (\ref{eq:hy}). Using
\begin{equation}
(h{\rm e}^{-\mu|\cdot|})(x)\le
\Big(\sum_{y\in\mbox{\boldmath $Z$}^d}|K_{0,y}|^s{\rm e}^{\mu|y|}\Big)
{\rm e}^{-\mu|x|}
\end{equation}
we see that $a\varphi(x)\le(h\varphi)(x)$ for $x\neq 0$
and hence
$\varphi(x)\le0$, i.e.,
\begin{equation}
g(x)\le g(0){\rm e}^{-\mu|x|}\;.
\label{eq:g}
\end{equation}
The claim follows now by combining this with
eq.~(\ref{eq:apb}).\hfill$\Box$

\medskip
For certain applications the following variant
of eq.~(\ref{eq:gs1}) is useful:

\begin{cor} Under the assumptions of
Theorem~\ref{thm:gs} we have
\begin{equation}
\mbox{\boldmath $E$}\Bigl(\Bigl|{G(x,y;z)\over G(x,x;z)}\Bigr|^s\Bigr)
\le {\rm e}^{-\mu |x-y|} \;,
\label{eq:gs2}
\end{equation}
for all $z\in\mbox{\boldmath $C$}\setminus\mbox{\boldmath
$R$}$.
\end{cor}

\medskip\noindent
{\bf Proof:} This is actually a corollary of the proof of
Theorem~\ref{thm:gs}. Due to
$\overline{G(x,y;z)}=G(y,x;\bar z)$ we may, upon interchanging $x$ and $y$,
prove (\ref{eq:gs2}) with $G(y,y;z)$ in the denominator. We then set $y=0$ as
before. A short computation based on eqs.~(\ref{eq:re}, \ref{eq:rk}) shows the
following dependence
\begin{equation}
{G(x,0;z)\over G(0,0;z)+{\rm i}\delta}={\alpha\over V_x-\beta}
\end{equation}
on $V_x$. If ${\rm Im}\,z>0$, as we may assume without loss, the
regularization by $\delta>0$ ensures that
\begin{equation}
|G(0,0;z)+{\rm i}\delta|^2=|G(0,0;z)|^2+2\delta{\rm Im}\,G(0,0;z)+\delta^2
\ge\delta^2\;,
\end{equation}
because of ${\rm Im\,}G(0,0;z)
={\rm Im\,}z\,\langle 0|(H-\bar z)^{-1}(H-z)^{-1}|0\rangle\ge 0$. We then
divide eq.~(\ref{eq:resolveqs}) by $|G(0,0;z)+{\rm i}\delta|^s$ and obtain
eq.~(\ref{eq:sh1}) once more but now for
\begin{equation}
g(x)=\mbox{\boldmath $E$}
\Bigl(\Bigl|{G(x,0;z)\over G(0,0;z)+{\rm i}\delta}\Bigr|^s\Bigr)\;.
\end{equation}
This is bounded in $x$ by $(\delta|{\rm Im}\,z|)^{-s}$. The upshot is
again eq.~(\ref{eq:g}) with $g(0)\le1$. Hence
\begin{equation}
\mbox{\boldmath $E$}
\Bigl(\Bigl|{G(x,0;z)\over G(0,0;z)+{\rm i}\delta}\Bigr|^s\Bigr)
\le {\rm e}^{-\mu |x-y|}\;,
\end{equation}
and the conclusion is by monotone convergence in the limit
$\delta\downarrow 0$. \hfill$\Box$


\maappendix{\bf Proof of the Decoupling Lemma}  \label{app:DL}

We shall need the inequality
\begin{equation}
|v-\beta|^{-s}+|u-\beta|^{-s}\le
{|v|^s\over|v-\beta|^s}(|u|^{-s}+|u-\beta|^{-s})+
{|u|^s\over|u-\beta|^s}(|v|^{-s}+|v-\beta|^{-s})
\label{eq:d1}
\end{equation}
for all $u,v,\beta\in\mbox{\boldmath $C$}$ (except for
vanishing denominators). Multiplication
by $|v-\beta|^s|u-\beta|^s$ shows it to be equivalent to
\begin{equation}
\Big({|v|^s\over|u|^s}-1\Big)|u-\beta|^s+|u|^s
+|v|^s+\Big({|u|^s\over|v|^s}-1\Big)|v-\beta|^s\ge0\;.
\label{eq:d2}
\end{equation}
Since this expression is symmetric in $u$ and $v$ it
suffices to prove this for $|u-\beta|\ge|v-\beta|$. The triangle
inequality yields
$|u-\beta|^s\le|v-\beta|^s+|u|^s+|v|^s$, which we apply to the two
middle terms of (\ref{eq:d2}) so as to bound it from below by
\begin{displaymath}
{|v|^s\over|u|^s}|u-\beta|^s+\Big({|u|^s\over|v|^s}-2\Big)|v-\beta|^s
\ge\Big({|v|^s\over|u|^s}+{|u|^s\over|v|^s}-2\Big)|v-\beta|^s\ge 0\;,
\end{displaymath}
since $t+t^{-1}\ge 2$ for $t>0$. This proves (\ref{eq:d1}).
Replace there $v$
by $v-\alpha$ and similarly for $u,\,\beta$, and integrate w.r.t.
$dq(u)dq(v)$. The result is
\begin{displaymath}
\int dq(u)\int dq(v){1\over|v-\beta|^s}\le
\int dq(v){|v-\alpha|^s\over|v-\beta|^s}
\int dq(u)(|u-\alpha|^{-s}+|u-\beta|^{-s})
\end{displaymath}
where, actually, each side comes duplicated with dummy
variables $u,\,v$
interchanged. The last integral is estimated by
(\ref{eq:in}).\hfill$\Box$

\maappendix{Analyticity and Exponential Decay
(proof of Theorem \ref{thm:FH})} \label{app:CT}

In section \ref{sec:2pt} we claimed and used the following statement.

\medskip\noindent{\bf Theorem \ref{thm:FH}\ } {\it
Let $H$ be as in eq.~(\ref{eq:h}) and $F$ be a function analytic
and bounded in the strip
$|{\rm Im}\,z|<\eta$ (by $ \|F\|_\infty$).
Then
\begin{equation}
|\langle x|F(H)|y\rangle|\le  18 \sqrt{2} \|F\|_\infty\
{\rm e}^{-\mu|x-y|}
\label{eq:appFH}
\end{equation}
for any $\mu$ such that the quantity
$b(\mu)=\sum_{x\in\mbox{\boldmath $Z$}^d}|K_{0,x}|$
$({\rm e}^{\mu|x|}-1)$ satisfies
$b(\mu) \le \eta/2$. }

By continuity it suffices to prove (\ref{eq:appFH})
in any smaller strip. We
may thus assume $F$ to be continuous up to the boundary.
We note that under
the above assumptions $F$ has the representation
\begin{eqnarray}
    F(E)&=&{1\over 2\pi{\rm i}}\int du
\Bigl[{1\over u-{\rm i}\eta}-
{1\over u+{\rm i}\eta}\Bigl]\, f(u-E)\nonumber\\
&=&D*f\,(E)
\label{eq:ct3}
\end{eqnarray}
with $D(u)= \eta / [(u^2+\eta^2) \pi]$
and $f$ a uniformly bounded function ($\|f\|_\infty<\infty$).
In fact, (\ref{eq:ct3}) is solved by
$f=F_++F_--D*F$, where
$F_\pm(u)=F(u\pm{\rm i}\eta\mp{\rm i}0)$.
This follows
from $D*(F_++F_-)=(D_++D_-)*F$ and $D_++D_-=\delta+D*D$.

The proof of Theorem~\ref{thm:FH} is related to the
Combes--Thomas bound \cite{CT}:
\begin{equation}
|G(x,y;E+{\rm i}\eta)| \le (2/\eta){\rm e}^{-\mu|x-y|}
\label{eq:ct}
\end{equation}
with $\mu$ as above.
In order to integrate over $u$ in \eq{eq:ct3} down to $-\infty$
we first develop the following related estimate.

\begin{lem}
With $\mu$ be small enough so that $b(\mu)\le\eta/2$,
\begin{eqnarray}
& |G(x,y;E+{\rm i}\eta)-G(x,y;E-{\rm i}\eta)| \ \le
12\ \eta {\rm e}^{-\mu|x-y|}
 &\nonumber \\
   &\cdot \langle x|{1\over (H-E)^2
+\eta^2/2}|x\rangle^{1/2} \langle y|{1\over(H-E)^2
+\eta^2/2}|y\rangle^{1/2}\;.
\label{eq:ctt}
\end{eqnarray}
\end{lem}
{\bf Proof:} We set $E=0$ for notational simplicity.
Let $2\pi\iu D=(H-\iu\eta)^{-1}-(H+\iu\eta)^{-1}$, and for any
bounded function $f(x)$ let
$D_f=\e^f D\e^{-f}+ \e^{-f} D\e^f$.  Since
\be
(\e^{f(x)-f(y)}+\e^{-(f(x)-f(y))})(G(x,y;\iu\eta)-G(x,y;-\iu\eta))
=2\pi\iu \langle x|D_f|y\rangle \;,
\label{eq:expg}
\ee
the desired bound would follow from the statement that
for any $f$ satisfying $|f(x)-f(y)|\le\mu|x-y|$ (e.g.,
a function which in a suitable finite region is $f(u)=\mu|u-y|$)
\be
\|[H^2 +\eta^2/2]^{1/2}D_f[H^2 +\eta^2/2]^{1/2}\|\le 6\eta/\pi
\label{eq:df} \; .
\ee

To prove eq.~(\ref{eq:df}), we first group the
terms as follows,
\bea
2\pi\iu D_f\,&=&{1\over H_f-\iu\eta}-{1\over H_f+\iu\eta}
        -{1\over H_f^*+\iu\eta}+{1\over H_f^*-\iu\eta}\nonumber\\
     &=&{1\over H_f-\iu\eta}B^+{1\over H_f^*+\iu\eta}
   -{1\over H_f+\iu\eta}B^-{1\over H_f^*-\iu\eta}\;,
   \label{eq:dd}
\eea
with $H_f=\e^f H \e^{-f}$, $B=H_f-H$,
and $B^\pm=(B^*-B\pm 2\iu\eta)$.
By the assumption on $\mu$, we have
\be
\|B\|\le b(\mu)\le\eta/2\;,\qquad
-3\eta\le\iu B^\pm\le 3\eta\;.
\label{eq:b}
\ee
We now claim that
\be
  (H_f^*-\iu\eta)(H_f+\iu\eta)\ge{1\over 2}[H^2+\eta^2/2]\;.
\label{eq:ct1}
\ee
Indeed, using the positivity of the last term in
\be
 (H_f^*-\iu\eta)(H_f+\iu\eta)
={1\over 2}[(H-\iu\eta)(H+\iu\eta)-2B^*B]
+{1\over 2}(H-\iu\eta+2B^*)(H+\iu\eta+2B)
\ee
and eq.~(\ref{eq:b}), we see that the l.h.s.
is bounded below by $(1/ 2)[H^2+\eta^2-(\eta^2/2)]$.

The estimates (\ref{eq:b}), (\ref{eq:ct1}), and equation
(\ref{eq:dd}), readily imply (\ref{eq:df}).
This bound, combined with an application of the Cauchy--Schwarz
inequality
to the r.h.s. of eq.~(\ref{eq:expg}) proves the claim made in
eq.~(\ref{eq:ctt}).  (The exponential weight is incorporated in
the l.h.s. in eq.~(\ref{eq:expg}).)

\medskip
\noindent {\bf Proof of Theorem~\ref{thm:FH}:}
Let us note that unlike the corresponding integral of
operator norms, the following integral is bounded:
\be
\int du \ \langle x| \frac{1}{(H-u)^2+\eta^2/2}
|x\rangle \   \le \sqrt{2}\ \pi / \eta
\label{eq:cauchy}
\ee
(using the spectral measure representation).
The claim, \eq{eq:appFH}, is obtained by combining
the integral representation of $F$, \eq{eq:ct3},
with the exponential bound \eq{eq:ctt}, and employing
the Cauchy--Schwarz inequality to reduce the resulting integral
to the one estimated in \eq{eq:cauchy}.
\hfill$\Box$

\maappendix{The ${\rm i}\eta$ Regularization\/}

The addition of a small imaginary term ${\rm i}
\eta$ to the energy is a standard regularization, and a convenient
alternative to the finite--volume cutoff.
Such a cutoff appears also in the Kubo formula for the
electrical conductivity in the absence of magnetic field, \eq{eq:co}.
Dealing with such expressions one should bear in mind the
operator bound: $\eta |G(0,x;E+{\rm i}\eta)| \le 1$.  For the
conductivity given by  \eq{eq:co} it implies that quite generally
\begin{equation}
\sigma_{i,j}(E) \le
\liminf_{\eta\downarrow 0}{\eta^s\over\pi} \sum_{x\in\mbox{\boldmath
$Z$}^d}|x_ix_j| \mbox{\boldmath $E$}(|G(0,x;E+{\rm i}\eta)|^s)
\;,
\label{eq:sgma}
\end{equation}
for any $s\le2$. Thus, the fractional moment localization
estimate, eq.~(\ref{eq:gs}), directly implies the vanishing
of the Kubo conductivity --- in the absence of magnetic field.

Working with this regularization, it is useful
to have also the following lemma.  Its second bound can
be used for yet another proof of the dynamical localization
(\ref{eq:exp1}), which
was originally derived using the finite--volume cutoff
(ref. \cite{Aiz94}) and which was provided another derivation
in ref. \cite{dRJLS}.

\begin{lem}\it If (\ref{eq:gs}) holds, and the
probability distribution, $q(dv)=r(v)dv$, satisfies the
regularity condition
$r\in L^1(\mbox{\boldmath $R$})\cap L^p(\mbox{\boldmath $R$})$
for some $p>1$,
then for any $E\in[a,b]$ and $\eta\neq0$,
\begin{eqnarray}
\eta^{1+p^{-1}}\mbox{\boldmath $E$}( |G(x,y;E+{\rm i}\eta)|^2 ) &\le
&C {\rm e}^{-\mu|x-y|}\;, \label{eq:p'}\\
\int_a^b dE\,\eta\mbox{\boldmath $E$}( |G(x,y;E+{\rm i}\eta)|^2 )
&\le &C {\rm e}^{-\mu|x-y|}\;. \label{eq:p}
\end{eqnarray}
\label{lm:ieta}
\end{lem}

To calibrate these statements we note that without any assumptions
on the self--adjoint operator $H$:
\begin{equation}
       \int_{-\infty}^{\infty} dE\,\eta |G(x,y;E+{\rm i}\eta)|^2  \le
\pi\;. \label{eq:p2}
\end{equation}

We shall only sketch here the proof of Lemma~\ref{lm:ieta},
which is by arguments seen in ref. \cite{Aiz94} (Lemma 3.1)
and in ref. \cite{Gf} (Lemma 3).
Some of the key points in the analysis are:\\
\noindent
i) Quite generally, for any $0\le s \le 2$:
\begin{equation}
|{\rm Im}\,z| |G(x,y;z)|^2\le
|{\rm Im}\,G(x,x;z)|
  \cdot{|G(x,y;z)|^s\over|G(x,x;z)|^s} \; .
\label{eq:p1}
\end{equation}
(For $s=0$ the proof is by a judicious use the
Cauchy--Schwarz inequality,  for $s=2$ it
follows from $|{\rm Im}\,z|\le|{\rm Im}\,G(x,x;z)^{-1}|$,
and for other $0\le s \le 2$ it holds by interpolation.) \\
\noindent
ii) Using eq. (\ref{eq:rk}) on the first factor on the
right side of (\ref{eq:p1})
yields
\ben
|\Im\, z| |G(x,y;z)|^2\le
{|\Im\, \widehat{G}(x,x;z)^{-1}|\over|V_x+\widehat{G}(x,x;z)^{-1}|^2}
  \cdot{|G(x,y;z)|^s\over|G(x,x;z)|^s}\;,
\een
where, again by (\ref{eq:rk}), the second quotient is
independent of $V_x$ (!).

To derive Lemma~\ref{lm:ieta}, one may now average first
over $V_x$ ---
in effect making use of the high degree of independence of
the values of the potential at different sites --- and then use
(\ref{eq:gs2}). The proof is most direct for the case of bounded density
$r(v)$ ($p=\infty$), and the extension to more singular distributions
is by arguments similar to those found in ref.~\cite{Aiz94}.

\medskip \noindent\remark
The condition expressed by the second statement in Lemma 4 implies
directly the exponential decay law
\be
    \E(\sup_{f:\|f\|_\infty \le1}|\langle x| P_{[a,b]}f(H)
|y\rangle|) \le C\e^{-\mu|x-y|}\;,
\ee
which is equivalent to eq.~(\ref{eq:exp1}). For that,
one may use the resolvents for an approximate $\delta$ function,
writing (for $f$ continuous):
$P_{[a,b]}f(H)=\mathop{\rm{s-}}\!\lim_{\eta\downarrow 0}f_\eta(H)$, where
\begin{equation}
f_\eta(H)
={1\over\pi}\int_a^b dE\,\eta\,{1 \over
H-E-{\rm i}\eta}{1 \over H-E+{\rm i}\eta}\,f(E)\;. \label{eq:feta}
\end{equation}
The matrix elements of (\ref{eq:feta}), can be easily
brought to a form in which (\ref{eq:p}) implies (\ref{eq:exp1}).
The key tool is
the Cauchy-Schwarz inequality, applied in the different setups:
the state Hilbert-space, $\mbox{\boldmath $E$}$ ---
the average over the disorder, and in $\int_a^b dE$.

\maappendix{Connes' area formula.}
\label{app:CF}

In Section~\ref{sec:SK}  an identity is stated relating
two expression for the Hall conductance: one based on the
charge-transport Index, and the other corresponding to the Streda
formula which takes the form of a Chern number, \eq{sigma3}.
Following is the derivation of Connes' area formula \cite{Connes}
which has been used to prove that relation.
The formulation and derivation presented here incorporate
a streamlined argument of Colin de Verdi\`ere (\cite{CdV,Av95}),
shown to us by R. Seiler.

\begin{thm}
For a fixed triplet
$u^{(1)},\,u^{(2)},\, u^{(3)}\in \mbox{\boldmath $Z$}^2$,
 let $\alpha_i(a)\in(-\pi,\,\pi)$ be the angle of view from
 $a\in \mbox{\boldmath $Z$}^{2*}$ of $u^{(i+2)}$ relative to $u^{(i+1)}$
  (with $\alpha_i(a)=0$ if $a$ lies between them).
 Let $g(\alpha)$ be an antisymmetric
 bounded function satisfying:
  \be
       g(\alpha) = \alpha +{\rm O}(\alpha^3)
  \label{eq:i5}
  \ee
  near $\alpha=0$. Then,
  \be
   \sum_{a\in\mbox{\boldmath $Z$}^{2*}}\sum_{i=1}^3 g(\alpha_i(a)) =
         2 \pi Area\left(\Delta (u^{(1)}, u^{(2)}, u^{(3)})\right)
  \label{eq:i6}
  \ee
where $ Area\left(\Delta (\ldots)\right)$ is the
triangle's {\em oriented area}.
  \end{thm}

Of special interest to us is the case with $g(\alpha) = \sin\alpha$,
which is used here in \eq{sigma3}.

\noindent{\bf Proof:}
We may assume the triangle to be positively oriented. The statement
(\ref{eq:i6}) is true
for $g(\alpha)=\alpha$. Indeed, for each $a\in \mbox{\boldmath $Z$}^{2*}$
\be
\sum_{i=1}^3 \alpha_i(a) = 2\pi
\left\{
\begin{array}{c}1 \\ 1/2 \\0\end{array}
\right\}
\hbox{for $a$}
\left\{
\begin{array}{c}
\hbox{inside}\\
\hbox{on the boundary of}\\
\hbox{outside}\\
\end{array}
\right\}
\hbox{the triangle}
\ee
Thus, for $g(\alpha) = \alpha$ the l.h.s. of (\ref{eq:i6}) is $2\pi
\times$ the
number of dual lattice sites within the triangle (counting a boundary
site with weight $1/2$).  This number is the same for triangles
obtained by the lattice translation and reflection symmetry
operations.    Since this set of triangles tiles the plane,
the number of enclosed dual sites must equal the triangle's area.

The above observation reduces \eq{eq:i6} to the statement  that for
$f(\alpha)=g(\alpha)-\alpha$
\be
\sum_{a\in\mbox{\boldmath $Z$}^2}\sum_{i=1}^3 \ f(\alpha_i(a))=0 \; .
\label{eq:F1}
\ee
A significant difference between $f$ and $g$ is that the individual
terms $f(\alpha_i(a))$
are summable in $a\in\mbox{\boldmath $Z$}^2$, since by  \eq{eq:i5}
$f(\alpha_i(a))={\rm O}(|a|^{-3})$ for $|a|\to\infty$.  However, each
of the three individual sums changes sign under the reflection with
respect to the midpoint of the corresponding edge,
$(a^{(i+1)}+a^{(i+2)})/2\in
(\mbox{\boldmath $Z$}/2)^2$ (which is a symmetry of the
 lattice $\mbox{\boldmath $Z$}^2$).
 Thus even the individual sums (at given $i$) vanish.
\hfill$\Box$

\medskip
\noindent {\it Acknowledgments.\/} We thank Y. Avron, J. Bellissard,
J. Fr\"ohlich, L. Pastur, R. Seiler, B. Simon and T. Spencer for valuable 
discussions. M.A. gratefully acknowledges the
hospitality accorded him at the Institute of Physics at Technion,
and at ETH-Z\"urich, and G.M.G. thanks T.~Spencer for an extended
stay at the Institute for Advanced Study.
The work was supported by the NSF Grant PHY-9512729.


\end{document}